\documentclass[iop,apjl]{emulateapj}
\newcommand{\V}[1]{\mathbf{#1}} %Bold for Vectors
\newcommand\Alfven{Alfv\'en } %Proper Names
\newcommand\Alfvenic{Alfv\'enic } %Proper Names
\newcommand\p[2]{\frac{\partial #1}{\partial #2}}
\usepackage{amssymb}
\usepackage{amsmath}
\usepackage{hyperref}
\usepackage{natbib}
\usepackage{color}
\usepackage{graphicx}% Include figure files
\usepackage{amsmath}
\usepackage{bm}

\begin{document}
\title{Evolution of 
The Proton Velocity Distribution due to Stochastic Heating
in the Near-Sun Solar Wind}

\author{Kristopher ~G. Klein \& Benjamin ~D.~G. Chandran}

\affiliation{Space Science Center, University of New Hampshire, 
Durham, NH 03824, USA}

\begin{abstract}
We investigate how the proton distribution function evolves
when the protons undergo stochastic heating by strong, low-frequency, Alfv\'en-wave
turbulence
under the assumption that $\beta$ is small.
We apply our analysis to protons undergoing stochastic heating in the
supersonic fast
solar wind and obtain proton distributions at heliocentric distances
ranging from 4 to 30 solar radii. 
We find that the proton distribution develops non-Gaussian structure
with a flat core and steep tail.
For $r >5 \ R_{\rm S}$,
the proton distribution is well approximated by
a modified Moyal distribution.
Comparisons with future measurements from \emph{Solar Probe Plus}
could be used to test whether stochastic heating is occurring
in the solar-wind acceleration region.
\end{abstract}


\maketitle
%-=-=-=-=-=-=-=-=-=-=-=-=-=-=-=-=-=-
%-=-=-=-=-=-=-=-=-=-=-=-=-=-=-=-=-=-
%=====================================================================

\section{Introduction}
\label{sec:intro}

Protons in the solar wind within a few solar radii of the Sun's
surface undergo an increase in temperature perpendicular to the
magnetic field $T_{p \perp}$
\citep{Kohl:1998,Esser:1999,Cranmer:1999}.  Additionally, as the solar
wind expands through the heliosphere, the total proton temperature
$T_p$ does not evolve as it would for a spherically expanding
adiabatic fluid, $T_p \propto r^{-4/3}$, but instead decreases more
slowly \citep{Hartle:1970}.  It is still unclear which mechanisms are
responsible for this observed behavior.

Many studies have proposed an injection of energy into the solar wind,
typically across some span of radial distance, 
with potential sources including
low-frequency \Alfvenic turbulence
\citep{Matthaeus:1999,Cranmer:2007}, 
ion cyclotron waves
\citep{Cranmer:2000,Hollweg:2002}, 
shocked compressive modes
\citep{Bruner:1978},
and magnetic reconnection, nano-flares,
and other impulsive events
\citep{Parker:1987,Cargill:2004,Drake:2009}.
Some authors have suggested that the dissipation of this energy 
is sufficient to accelerate the solar wind through the Sun's gravitational
well and then continue heating the solar wind as it progresses 
through the heliosphere
\citep{Matthaeus:1999,Smith:2001}. 
Other models contend that 
constructing the energy transport equations for
non-Gaussian velocity distributions, 
as are expected to arise in the nearly-collisionless
solar wind,
may be sufficient to explain radial temperature profiles
without the need for heating and acceleration from
energy dissipation \citep{Scudder:1992a,Scudder:2015}.
Identifying signatures of these proposed mechanisms that 
can be measured \emph{in situ} by spacecraft
is a necessary step for determining 
which mechanisms are responsible for governing the solar wind's evolution.

In this work, we focus on stochastic proton heating induced by strong,
low-frequency, Alfv\'en-wave turbulence.  A body of literature
concerning stochastic heating has shown that fluctuations with
amplitudes above a critical threshold are able to produce
perpendicular proton heating
\citep{McChesney:1987,Johnson:2001,Chen:2001,Voitenko:2004,Chaston:2004,Fiksel:2009}.
Because perpendicular ion heating is observed in the solar corona and
solar wind, and because Alfv\'en-wave turbulence is a dominant component of
solar-wind turbulence, stochastic proton heating has been proposed as a
candidate for explaining the observed coronal and solar-wind
temperature profiles.
\cite{Chandran:2010a} developed a phenomenological description of
the heating rates due to Alfv\'en-wave turbulence
with scale lengths comparable
to the ion gyroscale, and along with \cite{Chandran:2010b},
showed that such a mechanism could plausibly explain remote temperature 
observations in coronal holes. 
\cite{Chandran:2011} incorporated stochastic proton heating into 
a steady state two-fluid kinetic model of the fast solar wind,
finding that stochastic proton heating supplied a dominant contribution 
to the total turbulent heating rate.
\cite{vanderholst14} incorporated stochastic heating into 
three-dimensional, two-fluid, numerical simulations of the solar wind
and found promising agreement with observations.
\cite{Bourouaine:2013} found
that heating rates from \emph{Helios}
observations were consistent with stochastic proton heating occurring
in fast-wind streams between $0.29$ and $0.64$ AU.
\cite{Chandran:2013} showed that stochastic heating can explain
the observed alpha-particle-to-proton temperature ratio
\citep{Kasper:2013} and how this temperature ratio depends on 
the thermal-to-magnetic-pressure ratio $\beta$ 
and relative drift velocity $\Delta U_{\alpha p}$.

We extend these previous works by evaluating the effects of stochastic
heating on the evolution of the proton velocity distribution in a
model fast-wind stream in the inner heliosphere under the assumption
that $\beta \ll 1$.
We find that stochastic heating significantly alters the
distribution function from an (assumed) initial Gaussian to a flattop
distribution with steep tails.  These non-Gaussian features are
accurately described by a modified Moyal distribution.  The presence
or absence of this type of distribution function in low-$\beta$
regions could serve as a test for the importance of stochastic heating
in the solar wind.  We describe the equations of our model in
Section~\ref{sec:method} and present numerical solutions to these
equations in Section~\ref{sec:model}.

\section{Model of Stochastic Proton Heating in the Supersonic,
  Near-Sun Solar Wind}
\label{sec:method}

We model the evolution of the proton velocity distribution $f$
within a thin, open magnetic flux tube centered on a radially
oriented background magnetic field.  We start with the gyroaveraged kinetic
equation (\cite{Kulsrud:1983}'s Equation (37)) and drop the $\V{E}\times
\V{B}$ terms, leaving
\begin{equation}
\begin{aligned}
\p{f}{t} & + v_\parallel \V{\hat{b}} \cdot
\nabla f - \frac{1}{2}v_\perp v_\parallel \left(
\nabla \cdot \V{\hat{b}} \right) \p{f}{v_\perp}\\ &+\left( \frac{1}{2}v_\perp^2\nabla \cdot
  \V{\hat{b}} + \frac{q}{m} E_\parallel \right) \p{f}{v_\parallel} = 0,
\end{aligned}
\label{eqn:GA.1}
\end{equation}
where $q$ is the proton charge, $m$ the proton mass,
$\V{\hat{b}} = \V{B}/B$, $\V{B}$ is the magnetic field, and
$v_\parallel$ and $v_\perp$ are the proton velocity components
parallel and perpendicular to $\V{\hat{b}}$.  We assume that the
proton distribution function is gyrotropic.  The proton number density
$n$ and average radial flow velocity $U$ are given by
$n= \int d^3\V{v} f$ and $nU =\int d^3\V{v} v_\parallel f$.

We define the reduced distribution function
\begin{equation}
g(v_\perp) = 2 \pi \int_{-\infty}^\infty dv_\parallel f(v_\perp,v_\parallel).
\label{egn:g_def}
\end{equation}
We assume that the solar-wind outflow 
is supersonic and that, as a consequence, the average parallel proton
velocity is approximately $U$ at each $v_\perp$:
\begin{equation}
2\pi \int_{-\infty}^\infty dv_\parallel f(v_\perp,v_\parallel)
v_\parallel = Ug.
\label{eq:approx1} 
\end{equation} 
Upon integrating Equation~(\ref{eqn:GA.1}) over $v_\parallel$ and making
use of Equation~(\ref{eq:approx1}), we obtain an equation for the time
evolution of~$g$:
\begin{equation}
\begin{aligned}
\p{g}{t} & +  \V{\hat{b}} \cdot \nabla (Ug)  -
\frac{1}{2} v_\perp U \left(\nabla \cdot \V{\hat{b}}
\right)\p{g}{v_\perp} = 0.
\end{aligned}
\label{eqn:GA.2}
\end{equation}
We assume a steady state and that the magnetic field is nearly radial,
which enables us to rewrite Equation~\ref{eqn:GA.2} in the form
\begin{equation}
\frac{\partial }{\partial r} \left( Ug\right)
+ \frac{v_\perp U}{2B} 
\p{B}{r}
\p{g}{v_\perp} = 0.
\label{eqn:GA.3}
\end{equation}

To incorporate stochastic heating, we add a 
perpendicular-kinetic-energy diffusion term,
\begin{equation}
\begin{aligned}
\p{}{E_\perp}D_E\p{}{E_\perp}=
\frac{1}{v_\perp}\p{}{v_\perp}\frac{D_E}{m^2 v_\perp}\p{}{v_\perp},
\end{aligned}
\label{eqn:DE}
\end{equation}
to the right-hand side of Equation~\ref{eqn:GA.3}:
\begin{equation}
\begin{aligned}
\frac{\partial}{\partial r} (Ug)
+ \frac{v_\perp U}{2B} \p{B}{r}
\p{g}{v_\perp} = \frac{1}{v_\perp}\p{}{v_\perp}\frac{D_E}{m^2
  v_\perp}\p{g}{v_\perp}.
\end{aligned}
\label{eqn:GA.4}
\end{equation}
For the diffusion coefficient $D_E$, we use the expression
in Equation 17 from \cite{Chandran:2010a}
multiplied by an exponential suppression term,
\begin{equation}
\frac{D_E}{m^2 v_\perp}
\sim \delta v^3_\rho \Omega_p 
\exp \left[-\frac{c_2}{\epsilon}\right],
\label{eqn:DE.3}
\end{equation}
where $\delta v_\rho$ is the 
rms amplitude of the $\bm{E}\times\bm{B}$ velocity fluctuations at scale
$\rho = v_\perp /\Omega_p$, and $\Omega_p = q B_0/m c$ is the proton
gyrofrequency.  The effectiveness of stochastic heating depends
strongly on the stochasticity parameter
$\epsilon \equiv \delta v_\rho/v_\perp$.  When $\epsilon \ll 1$, a
proton's gyromotion is only weakly perturbed by the gyroscale
fluctuations and the proton's magnetic moment is nearly conserved.
When $\epsilon \sim 1$, the gyroscale electric-field fluctuations
strongly distort a proton's gyromotion, which enables the time-varying
electrostatic potential to cause perpendicular proton heating.
\cite{Chandran:2010a} incorporated the exponential suppression term
$\exp \left( -c_2 /\epsilon_p \right)$ into their expression for the
stochastic heating rate of the entire distribution, where
$\epsilon_p = \epsilon|_{v_\perp = w_p}$ and
$w_p = (2 k_B T_\perp/ m)^{1/2}$.  In Equation~\ref{eqn:DE.3}, we
incorporate an analogous but velocity-dependent exponential
suppression term directly into the velocity-diffusion coefficient.
The value of $c_2$, taken to be $0.2$ in this work, modifies the
amplitude threshold above which stochastic heating becomes effective.
This choice of $c_2$ is motivated by
values found in test-particle simulations \citep{Xia:2013} and
inferred from \emph{Helios 2} observations \citep{Bourouaine:2013}.
We note that $D_E$ was derived in the low-$\beta$ limit, and therefore
the model constructed below is also restricted to solar-wind streams
with small $\beta$.

\begin{figure}[t]
\includegraphics[width=8.5cm, viewport = 0 0 185 170, clip = true]
%{f1.eps}
{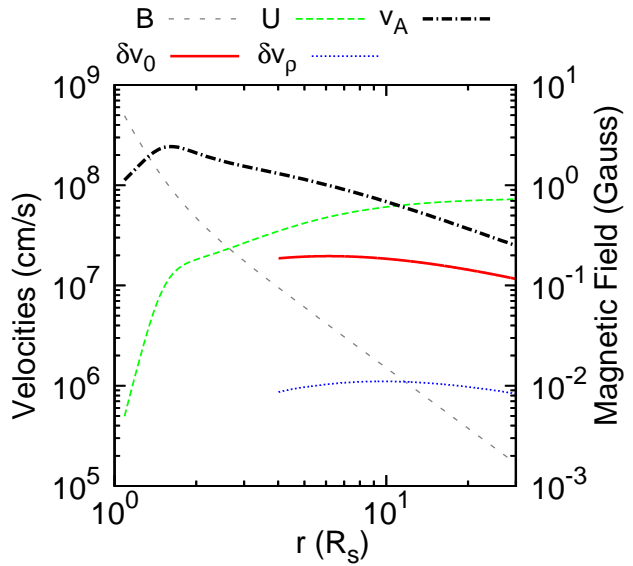}
\caption{ Input radial profiles for velocity (left axis) and magnetic
  field (right axis) amplitudes.  The solar-wind (black) and \Alfven
  (green) velocities as well as the magnetic field  (grey) are given in
  Equations~\ref{eqn:B}-\ref{eqn:U}.  RMS velocity-fluctuation
  amplitudes at the outer scale and thermal gyroscale from
  Equations~\ref{eqn:dvp} and \ref{eqn:dv0} are also plotted.  }
\label{fig:UB}
\end{figure}

To solve Equation~\ref{eqn:GA.4}, we first specify radial
profiles for the magnetic field $B_0$, solar-wind velocity $U$,
and diffusion coefficient $D_E(v_\perp,r)$.
We adopt the magnetic-field profile used by 
\cite{Hollweg:2002},
\begin{equation}
B_0 = \left[
\frac{6}{x^6} + \frac{1.5}{x^2}
\right]
{\rm Gauss},
\label{eqn:B}
\end{equation}
where
\begin{equation}
x = \frac{r}{R_{\rm s}},
\label{eq:defx} 
\end{equation} 
and the density profile of \cite{Feldman:1997}
with the $x^{-2}$ term added by \cite{Chandran:2009d},
\begin{equation}
n = \left(
 \frac{3.23 \times 10^8}{x^{15.6}}
+\frac{2.51 \times 10^6}{x^{3.76}}
+\frac{1.85 \times 10^5}{x^2}
\right)
{\rm cm^{-3}}.
\label{eqn:n}
\end{equation}
Upon multiplying Equation~(\ref{eqn:GA.4}) by $v_\perp$ and
integrating over~$v_\perp$, we obtain
\begin{equation}
\frac{d}{dr} \left( \frac{nU}{B_0}\right) = 0,
\label{eq:cons}
\end{equation} 
which can also be deduced by simply noting that the magnetic flux~$B_0A$
and ``proton flux'' $n U A$ through a narrow flow/flux tube of cross sectional
area~$A$ are both independent of~$r$ in steady state.
From Equation~(\ref{eq:cons}) we obtain
\begin{equation}
U = 9.25 \times 10^{12} \ \frac{B_0}{n} \
{\rm cm \ s^{-1}},
\label{eqn:U}
\end{equation}
where the numerical coefficient has been chosen so that $U$
extrapolates to a value of 750~km/s at 1~AU.
Profiles for $B_0$ and $U$ are plotted in 
Figure~\ref{fig:UB}, and $n$ is plotted in panel (a) of Figure~\ref{fig:moments}.

We define $\delta v_\lambda$ to be the rms amplitude of the
$\bm{E}\times\bm{B}$ velocity at a scale length~$\lambda$ 
measured perpendicular to~$\bm{B}_0$.
Based on theoretical predictions
\citep{Boldyrev:2006,Chandran:2015}, numerical simulations
\citep{Boldyrev:2011,Perez:2012}, and solar-wind observations
\citep{Podesta:2009}, we set
\begin{equation}
\delta v_\lambda = \delta v_0 \left(\frac{\lambda}{L_0}\right)^{1/4}
\label{eqn:dvlambda}
\end{equation}
when $\lambda$ is in the inertial range, where $L_0$ is
the outer-scale correlation length.
We take
$L_0$ to be proportional to the radius 
of the flux tube in which the turbulence is embedded.
Specifically, we use Equation (33) from \cite{Chandran:2009d}, setting
$L_0 =5000 [7.5 \ {\rm G} /B_0(r)]^{1/2} \ {\rm km}$.

We assume that Equation~(\ref{eqn:dvlambda}) can be extrapolated, at
least approximately,  all the way to the
proton-gyroradius scale~$\rho$, which yields 
\begin{equation}
\delta v_\rho = \delta v_0 \left(\frac{\rho}{L_0}\right)^{1/4}.
\label{eqn:dvp}
\end{equation}
This assumption neglects the possible
  back reaction of the heating process on the turbulent power
  spectrum. For example, if stochastic heating drains energy from
  the turbulent cascade at scales~$\sim \rho$, it could reduce $\delta v_\rho $ below the
  value in Equation~(\ref{eqn:dvp}). \cite{Chandran:2010b} modeled
  this back reaction by including a reduction factor in the expression
  for~$\delta v_\rho$. In the present paper, we find that even without
 such a reduction factor the ion heating rate remains less than the cascade power except
  within a narrow radial interval near the innermost radius~$r_0$ in our
  numerical solutions (see Equation~(\ref{eq:Qturb}) and
  Figure~\ref{fig:heating}), which is dominated by artifacts
  associated with our imposition of a  Gaussian velocity distribution at~$r_0$
  (see discussion below). We have thus refrained from multiplying the
  right-hand side of Equation~(\ref{eqn:dvp}) by a reduction factor in
  order to keep our model as simple as possible.

For the radial profile of $\delta v_0$, 
we use the reflection-driven turbulence
model of \cite{Chandran:2009d},
converting their expression for the 
outward-propagating Heinemann-Olbert variable $g_{\rm HO}$ into
the velocity amplitude
\begin{equation}
\begin{aligned}
\delta v_0 & = \frac{g_a}{2}\left(\frac{\eta^{1/4}}{1+ \eta^{1/2}}\right)
    \left(\frac{v_A}{v_{Aa}}\right)^{\chi/2},
\end{aligned}
\label{eqn:dv0}
\end{equation}
where $v_A = B_0/(4 \pi m n)^{1/2}$ is the \Alfven speed, $\eta
\equiv n/n_a$, and $v_{Aa}$, $g_a$, and $n_a$ are values of
$v_A$, $g_{\rm HO}$, and $n$ at the \Alfven critical point, which is
at $r=11.1 \ R_{\rm S}$ in our model.  The factor $\chi$ models
the reduction of the efficiency of wave reflection for waves with
periods below $\sim 1$ hour.  Both values of $\chi$ used in
\cite{Chandran:2009d}, $\chi =1$ and $0.65$, lead to qualitatively
similar heating rates and velocity distributions in our model.  In
this work, we only present results from the $\chi = 0.65$ model, which
gives turbulence amplitudes closer to observed values.  The value of
$g_a$ is set to $7.2 \times 10^{7} {\rm cm \ s^{-1}}$ to match
constraints from \emph{Helios 2} \citep{Marsch:1982} and 
\emph{Ultraviolet Coronagraph Spectrometer} (\emph{UVCS})
observations \citep{Esser:1999}. Profiles for $\delta v_0$ and
$\delta v_\rho$ are shown in Figure~\ref{fig:UB}.
With the above assumptions, we can rewrite Equation~(\ref{eqn:DE.3}) 
as
\begin{equation} 
\frac{D_E}{m^2 v_\perp} =  \Omega_p^{1/4}  \delta v_0^3 
\left(\frac{v_\perp}{  L_0 } \right)^{3/4} 
\exp \left(
\frac{-c_2 v_\perp^{3/4} \left(\Omega_p  L_0 \right)^{1/4}}
{\delta v_0 }
\right).
\label{eqn:DE.4}
\end{equation}

\section{Numerical Results}
\label{sec:model}

We solve Equation~\ref{eqn:GA.4} numerically using the Crank-Nicholson
finite-difference method.  The reduced distribution function
$g(v_\perp, r)$ is evaluated at $10^5$ points logarithmically spaced
between $r_0=4 \ R_{\rm S}$ and $r_{\rm f} = 30 \ R_{\rm S}$.  The
value of $r_0$ is chosen so that the solar wind is supersonic and
weakly collisional. 
Based upon \emph{UVCS} observations \citep{Kohl:1998,Esser:1999}, we take
$T_{\perp0}$ to be $2 \times 10^6 \ {\rm K}$.  We also take
$g(v_\perp, r_0)$ to be a Gaussian.  
The initial density,
$n_0 = 2.51 \times 10^{4} \ {\rm cm^{-3}}$, is found through
evaluation of Equation~\ref{eqn:n} at $r_0$.

The velocity grid has $10^3$ linearly spaced points between
$v_{\perp,0} = 0$ and $v_{\rm \perp,f}=10 w_{0}$, where $w_{0} = 1.82
\times 10^7 \ {\rm cm \ s^{-1}}$ is the thermal speed of the Gaussian
distribution at $r_0$.  We set $\partial g/\partial v_\perp= 0 $ at
$v_\perp = 0$ and $g(v_{\rm \perp,f})= 0$ at $v_\perp = v_{\rm \perp,
  f}$.  The choice of velocity and radial resolutions is sufficient to
remove the spurious decaying oscillations that arise for
under-resolved Crank-Nicholson solutions.  To check that our outer
boundary condition does not significantly alter our results, we
recalculate our numerical solution with $v_{\rm \perp, f}=15 w_{0}$
and $20 w_{0}$, retaining the same resolution in $v_\perp$. These
extended models (not shown) produce virtually the same reduced distribution
functions as the model with $v_{\rm \perp, f}=10 w_{0}$.  We have
verified that our numerical method conserves particles. 

\begin{figure}[t]
\includegraphics[width=8.5cm, viewport = 30 15 320 145, clip = true]
%{f2.eps}
{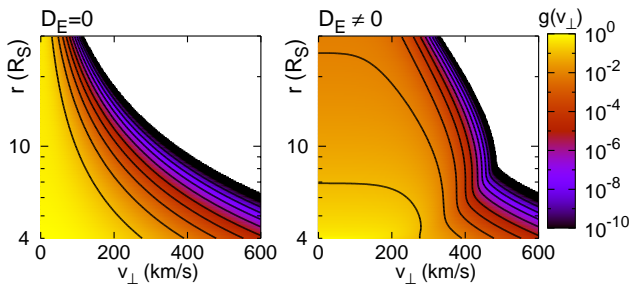}
\caption{
Contour plots of the reduced distribution function $g(v_\perp)$ as a function
of radial distance for the $D_E = 0$ (left) and $D_E \neq 0$ (right) cases.
Contour lines indicate increments of factors of 10.
}
\label{fig:g_contour}
\end{figure}

We solve
Equation~\ref{eqn:GA.4} with $D_E$ given by Equation~(\ref{eqn:DE.4}) 
and with $D_E$ set equal to zero for comparison.
We present contour plots of $g(v_\perp, r)$ in
Figure~\ref{fig:g_contour} for the $D_E =0$ (left panel) and
$D_E \neq 0$ (right) cases.  When $D_E = 0$, particles shift from high
to low $v_\perp$ as a result of magnetic moment conservation.  When
$D_E$ is given by Equation~(\ref{eqn:DE.4}), the narrowing of $g$ is
arrested and reversed between $7$ and $10 \ R_{\rm S}$.  At larger
distances, particles do lose energy as they flow away from the Sun,
but at a much slower rate than in the $D_E =0$ case.

\begin{figure*}[t]
\includegraphics[width=17.5cm, viewport = 0 0 475 150, clip = true]
%{f3.eps}
{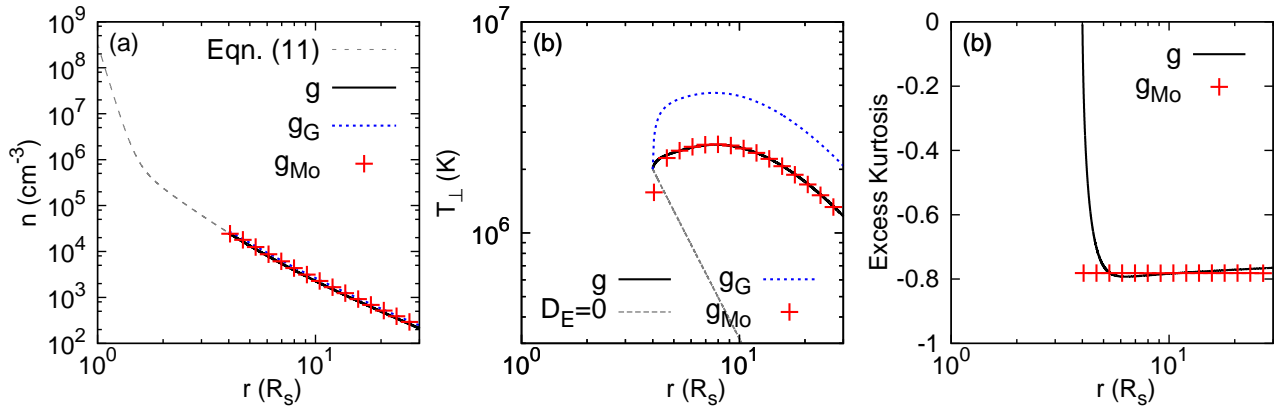}
\caption{Radial profiles  of $n$, $T_\perp$, and $\kappa$ of the reduced distribution function $g$ (black) as
  well as the best-fit Gaussian $g_{\rm G}$ (blue) and modified Moyal $g_{\rm Mo}$
  (red) distributions.  (a) Density profiles from $g$ match the
  input profile (Equation \ref{eqn:n}, grey dashed) as do the fitted
  densities $n_{\rm Mo}$ and $n_{\rm G}$.  (b) Temperature profiles
  for $D_E \neq 0 $ (black) show an increasing then slowly decreasing
  $T_\perp$, compared to the $D_E=0$ case (grey dashed), in which
  $T_\perp$ decreases rapidly with increasing~$r$.  The best-fit Gaussian $g_{\rm G}$
  systematically overestimates $T_\perp$, but $T_{\rm Mo, \perp}$ is
  in good agreement with $T_\perp$ beyond $5 R_{\rm S}$. (c) The excess kurtosis is 
  less than the Gaussian value of zero and in good agreement with
  $\kappa_{\rm Mo}=-0.781$.}
\label{fig:moments}
\end{figure*}

In Figure~\ref{fig:moments}, we plot radial profiles of 
the density 
\begin{equation}
n(r) = \int_0^{\rm v_\perp, f} dv_\perp v_\perp g(v_\perp, r),
\label{eqn:nn}
\end{equation}
perpendicular temperature
\begin{equation}
T_\perp(r) = \frac{m}{2 k_B n(r)}\int_0^{\rm v_\perp, f} dv_\perp v_\perp^3 
g(v_\perp, r),
\label{eqn:Tperp}
\end{equation}
and excess kurtosis
\begin{equation}
\kappa (r) = \frac{n(r) \int_0^{\rm v_\perp, f} dv_\perp v_\perp^5 g(v_\perp, r)}
{\left(\int_0^{\rm v_\perp, f} dv_\perp v_\perp^3 g(v_\perp, r)\right)^2} - 3.
\label{eqn:kurtosis}
\end{equation}
The density profiles for the $D_E=0$ and $D_E \neq 0$ cases are the same
and in agreement with the input density profile, Equation~(\ref{eqn:n}),
as they must be since Equation~(\ref{eqn:n}) follows from
Equations~(\ref{eqn:GA.4}), (\ref{eqn:B}), and (\ref{eqn:U})  via
Equation~(\ref{eq:cons}). 
The perpendicular temperature $T_\perp$ initially decreases, before
increasing to a peak of $2.5 \times 10^6 \ {\rm K}$ at $8 \ R_S$.  The
temperature then falls monotonically to $1.2 \times 10^6 \ {\rm K}$ at
$30 \ R_S$.  As foreshadowed by the left panel of
Figure~\ref{fig:g_contour}, $T_\perp$ for the $D_E=0$ case
rapidly falls from $r_0$ outward, reaching $6.6\times 10^3
\ {\rm K}$ at $30 \ R_{\rm S}$.

A Gaussian distribution has no excess kurtosis: $\kappa_{\rm G}= 0$.
Our $D_E \neq 0$ distribution is platykurtic, with negative excess kurtosis.
For the reduced distribution, 
$\kappa$ departs from $0$ rapidly before leveling off at
$\kappa \approx -0.8$ near $5 \ R_{\rm S}$.
The value of $\kappa$ then gradually increases for larger radial distances, but the
distribution remains highly non-Gaussian over the entire radial range under 
consideration in this work.

\begin{figure*}[t]
\includegraphics[width=17.5cm, viewport = 0 0 475 150, clip = true]
{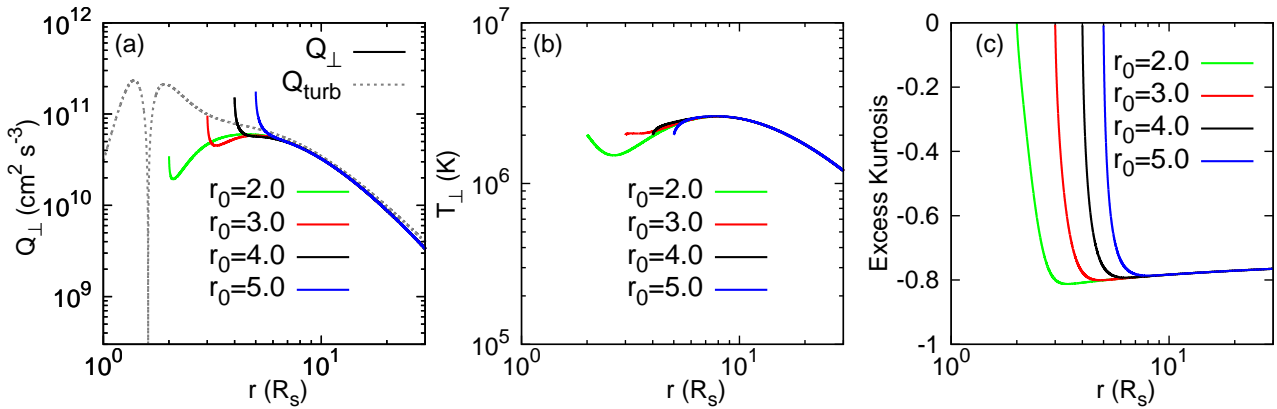}
\caption{Radial profiles of $Q_\perp$, $T_\perp$, and $\kappa$ for
  $r_0$ varying from $2$ to $5 \ R_{\rm S}$.  In panel (a), $Q_\perp$
  from Equation~\ref{eqn:Qperp} (black line for $r_0 = 4 \ R_{\rm S}$)
  is compared to $Q_{\rm turb}$ from the model of
  \cite{Chandran:2009d} (grey dashed). Across all three panels,
  quantities converge to the same radial profile within one to two
  solar radii of $r_0$.}
\label{fig:heating}
\end{figure*}

We calculate the perpendicular heating rate per unit mass, $Q_\perp$,
by multiplying Equation~\ref{eqn:GA.4} by $v_\perp^3/2n$
and integrating over $v_\perp$, which gives
\begin{equation}
 \frac{B  k_B U}{m} \frac{d}{dr} \left(
\frac{T_\perp}{B}\right) = Q_\perp,
\label{eqn:2ndMom}
\end{equation}
where 
\begin{equation}
Q_\perp =  \frac{1}{2n} \int_0^{\infty} d
v_\perp v_\perp^2 \p{}{v_\perp}\frac{D_E}{m^2 v_\perp}\p{g}{v_\perp}.
\label{eqn:Qperp}
\end{equation}
We plot $Q_\perp$ in Figure~\ref{fig:heating}. For this figure, we
approximate
Equation~(\ref{eqn:Qperp}) by replacing the upper limit of integration
with $v_{\perp, \rm f}$.

Beyond the immediate neighborhood of $r_0$, as $r$ increases $Q_\perp$ levels off
near $5 \ R_{\rm S}$ and then decreases monotonically.  
Our assumption of a Gaussian distribution at $r_0$ is for concreteness.
We do not expect this assumption to hold in the corona, and therefore
our model is inaccurate near $r_0$.
To understand the effects of the boundary condition at $r_0$, we
recalculate the numerical solution using $r_0 \in [2,3,5]
\ R_S$, keeping $T_{\perp 0} = 2\times 10^6 \ {\rm K}$, and using
$n_0(r_0)$ from Equation~\ref{eqn:n}.  While the resulting heating rates
differ near $r_0$, all of the solutions quickly converge to the
same radial profile.  A similar convergence
is seen for $T_\perp$ and $\kappa$ in Figure~\ref{fig:heating}, as well as for
the reduced distribution function $g$ (not shown).

We 
compare the perpendicular proton heating rate $Q_\perp$ with the
approximate total turbulent heating rate (per unit mass)
\begin{equation}
Q_{\rm turb} =
\frac{z^-(z^+)^2}{4 L_0}
\label{eq:Qturb} 
\end{equation}
in the model of \cite{Chandran:2009d}, where $z^+$ ($z^-$) is the rms
amplitude of the Alfv\'en-wave-like fluctuations that are outward-propagating (inward-propagating) 
when viewed in the local plasma frame.
Except in the immediate vicinity of $r_0$, the
perpendicular heating rate is less than the turbulent heating rate,
indicating that not all of the cascade power is dissipated
by protons via stochastic heating.

To characterize the shape of the velocity distribution, we consider two
types of fits to~$g$. First,
we fit $g$ using a least-squares method to a Gaussian
of the form
\begin{equation}
g_{\rm G}(v_\perp,r) = \frac{2 n_{\rm G}(r)}{w_{\rm G}^2(r)}
 \exp \left(-\frac{v_\perp^2}{w_{\rm G}^2(r)}\right)
\label{eqn:gauss}
\end{equation}
at each radial grid point.
The density and perpendicular temperature
($T_{\rm G,\perp} = m w_{\rm G}^2 / 2 k_B$) 
profiles calculated from the fitted $g_{\rm G}$ are shown 
in panels (a) and (b) of Figure~\ref{fig:moments}.
The Gaussian density $n_{\rm G}$ agrees with the input density,
while $T_{\rm G,\perp}$ overestimates $T_\perp$ by a factor that is
$\simeq 1.7$ at $r\gtrsim 5 R_{\rm s}$.

\begin{figure}[t]
\includegraphics[width=8.5cm, viewport = 0 5 350 145, clip = true]
{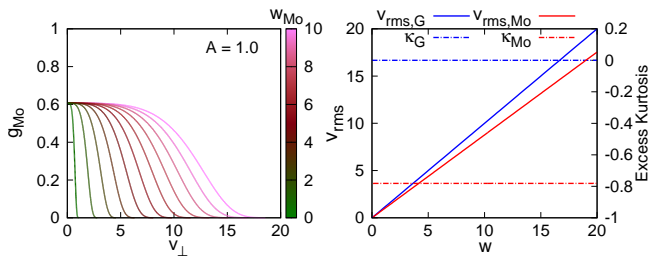}
\caption{
Properties of the modified Moyal distribution.
Panel (a): $g_{\rm Mo}(v_\perp)$ for $A = 1.0$
and for $w_{\rm Mo} $ ranging from $0.5$ (green) to $10$ (pink). 
Panel (b): $v_{\rm rms}$ (solid lines; left axis) and
$\kappa$ (dashed-dotted; right axis)
for $g_{\rm G}$ (blue) and $g_{\rm Mo}$ (red)
as a function of thermal speed $w$.
}
\label{fig:g_M}
\end{figure}

We next fit $g$ to a modified Moyal distribution \citep{Moyal:1955}
of the form
\begin{equation}
g_{\rm Mo}=A(r) 
\exp\left( \frac{1}{2}\left[
\frac{v_\perp^2}{w_{\rm Mo}^2(r)}
- \exp \left( \frac{v_\perp^2}{ w_{\rm Mo}^2(r)}\right)
\right]
\right).
\label{eqn:moyal}
\end{equation}
As this
distribution is not frequently employed in the solar-physics literature,
we plot selected properties in Figure~\ref{fig:g_M}.  In the left panel,
$g_{\rm Mo}$ is plotted with fixed $A=1.0$ for $w_{\rm Mo}$ varying
from $0.5$ to $10$.  In the right panel, the root mean square velocity and
excess kurtosis for both $g_{\rm G}$ and $g_{\rm Mo}$ are plotted as a
function of $w$.  In the Gaussian case, $v_{\rm rms} = w_{\rm G}$,
while for the modified Moyal distribution, $v_{\rm rms} \approx 0.877
w_{\rm Mo}$.  Both distributions have a constant $\kappa$, $\kappa_{\rm
  G}=0$ and $\kappa_{\rm Mo} \approx -0.781$.

At each radial grid point, we calculate the best-fit modified Moyal
distribution $g_{\rm Mo}(v_\perp,r)$ for the reduced distribution
$g(v_\perp, r)$.  We then calculate the density, perpendicular
temperature ($T_{\rm Mo,\perp} = m v_{\rm rms}^2/2 k_B$), and excess
kurtosis $\kappa_{\rm Mo}$ by numerically integrating the fitted
function and plot these quantities in panels (a)-(c) of
Figure~\ref{fig:moments}.  The fitted Moyal density in panel (a) is in
agreement with the input density, and overlaps with plots of $n$.  The
temperature corresponding to the best-fit Moyal distribution,
$T_{\rm Mo,\perp}$ (panel b), slightly underestimates $T_\perp$
between $r_0$ and $5 \ R_{\rm S}$, beyond which $T_{\rm Mo,\perp}$ and
$T_\perp$ are in good agreement.  The fitted excess kurtosis
$\kappa_{\rm Mo}$ has a constant value of $-0.781$ for all radial
distances, which compares well to $\kappa$ calculated from $g$.

\begin{figure*}[ht]
\includegraphics[width=18.5cm, viewport = 0 0 380 160, clip = true]
%{f6.eps}
{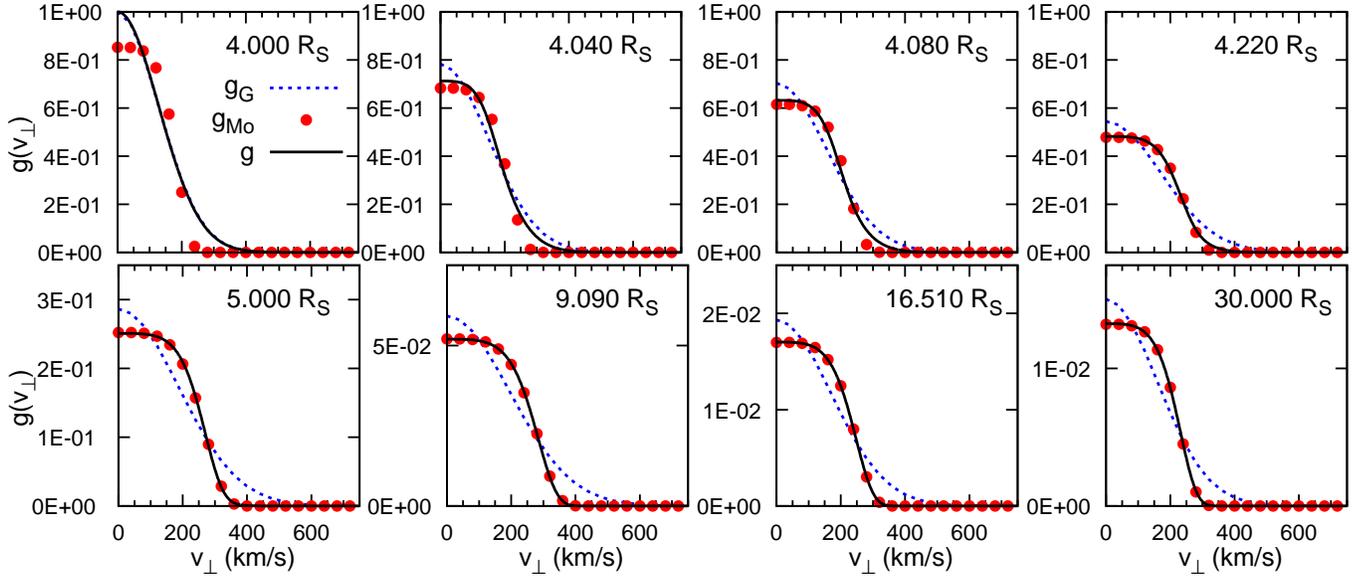}
\caption{
The reduced distributions $g$ (black lines)
and best fits $g_{\rm G}$ (blue lines) and 
$g_{\rm Mo}$ (red dots) 
at eight radial distances.
Note the change in the y-axis for each plot, which is necessitated by
the dramatic drop in density over the radial distances investigated;
the x-axis is kept constant over the eight plots.
}
\label{fig:g_slice}
\end{figure*}

In Figure~\ref{fig:g_slice} we plot $g$ and the associated fit
functions $g_{\rm G}$ and $g_{\rm Mo}$ at eight radial
distances. Near $r_0 = 4 \ R_{\rm S}$,  we see that the
core of $g$ has flattened, but that the tail is still Gaussian.
By $5 \ R_{\rm S}$, the Gaussian features of $g$ have disappeared,
and $g_{\rm Mo}$ is visually indistinguishable from $g$.  When
compared to the Gaussian $g_{\rm G}$, $g$ has a much flatter core and
a significantly steeper tail.  This shape is a product of the rapid
energy diffusion at small $v_\perp$ and negligible energy diffusion at
superthermal perpendicular velocities. 

\section{Conclusion}
\label{sec:discussion}

We have solved for the evolution of the reduced distribution function
$g(v_\perp)$ in a model fast-solar-wind stream in the presence of
stochastic heating under the assumption that $\beta$ is small.
Stochastic heating produces perpendicular proton heating, as is
observed in the solar wind.  It also causes $g$ to develop significant
non-Gaussian features, specifically a platykurtic (negative excess
kurtosis) flattop distribution, which is well modeled by a modified
Moyal distribution (Equation~\ref{eqn:moyal}).
Detailed measurements of the proton distribution function from
the upcoming \emph{Solar Probe Plus} mission
will provide a wealth of
data that can be compared with the model results presented here,
providing a test for the importance of stochastic heating in the 
solar-wind acceleration region.

\acknowledgements We thank Phil Isenberg for helpful discussions.
This work was supported by NSF grants AGS-1258998, AGS-1331355, and
PHY-1500041 and NASA grant NNX15AI80G.

%-=-=-=-=-=-=-=-=-=-=-=-=-=-=-=                                                 
\bibliographystyle{apj}

%\bibliography{stochastic_diff}

%-=-=-=-=-=-=-=-=-=-=-=-=-=-=-=    

\end{document}